\documentclass[ 
superscriptaddress,
amsmath,amssymb,
aps,
prb,
prl,
reprint,
]{revtex4-2}

\usepackage{graphicx}
\usepackage{dcolumn}
\usepackage{bm}
\usepackage{mathtools}
\usepackage{braket}
\usepackage[bb=boondox]{mathalfa} 

\usepackage{amsmath}

\usepackage{mathrsfs}  
\usepackage{graphicx} 
\usepackage{caption}
\usepackage{float}
\usepackage{color}
\usepackage{xcolor}
\definecolor{red}{rgb}{0,0,0} 

\usepackage{booktabs} 
\usepackage{array} 
\usepackage{paralist} 
\usepackage{verbatim} 
\usepackage{subfig} 
\usepackage{float} 
\usepackage{hyperref}
\hypersetup{colorlinks=true,linkcolor=blue,urlcolor=blue}

\usepackage{xcolor}
\newif\ifproofread

\begin{document}

\preprint{APS/123-QED}

\title{{\color{black} Multi-mode Floquet NEGF method for driven quantum transport}}

\author{Vahid Mosallanejad}
\email{vahid@westlake.edu.cn} 
\affiliation{Department of Chemistry, School of Science, Westlake University, Hangzhou, Zhejiang 310024, China}
\affiliation{Institute of Natural Sciences, Westlake Institute for Advanced Study, Hangzhou, Zhejiang 310024, China}


\author{Wenjie Dou}
\email{douwenjie@westlake.edu.cn} 
\affiliation{Department of Chemistry, School of Science, Westlake University, Hangzhou, Zhejiang 310024, China}
\affiliation{Department of Physics, School of Science, Westlake University, Hangzhou, Zhejiang 310024, China}
\affiliation{Institute of Natural Sciences, Westlake Institute for Advanced Study, Hangzhou, Zhejiang 310024, China}

\date{\today}
\begin{abstract}
We present a non-perturbative Floquet-based non-equilibrium Green's function (NEGF) method to study electron transport in a quantum system driven simultaneously by multiple independent terms (multi-mode). We first derive the two-mode Floquet NEGF based on two-step transformations of the retarded-advanced Green's function from the Kadanoff–Baym equation. This derivation proceeds by elaborating on the expectation values of the number and current operators. The two-mode Floquet NEGF is then extended to cases with multiple drivings. The method is tested by investigating current suppression in the presence of two drivings. We show that an extra sinusoidal off-diagonal driving can cause substantial modification to the current suppression, provided careful selection of the driving frequency. Consequently, we expect that the established method has broad applications in a wide range of open quantum systems driven by complicated drivings.
\end{abstract}

\maketitle
For close quantum systems, applying a strong drive gives rise to interesting phenomena, such as the Landau-Zener-St\"uckelberg (LZS) interference \cite{landau1932theorie, zener1932non, stuckelberg1932theorie, higuchi2017light,heide2018coherent,ota2018landau}, ac Stark effect~\cite{autler1955stark, xu2008single,schneider2018local,garzon2018stark}, multi-photon processes~\cite{hernandez2013attosecond,shcherbakov2021generation,schmid2021tunable}, dynamical localization/coherent destruction of tunneling (CDT)~\cite{dunlap1986dynamic,grossmann1991coherent,grossmann1992localization, kayanuma2008coherent}, inverse Faraday effect~\cite{van1965optically,battiato2014quantum,mironov2021inverse}, etc. Such effects have real world applications. For example, LZS interference provides means for quantum computing control on qubits~\cite{stehlik2012landau,cao2013ultrafast,wang2018landau}. 
For systems open to exchange charge carriers with terminal (bath), the key signatures of driven transport are inducing; Photon-assisted tunneling~\cite{kouwenhoven1994observation,kouwenhoven1994photon,keay1995photon,platero2004photon, hazelzet2001coherent}, pump current~\cite{kouwenhoven1991quantized, feve2007demand,wang2022generation, restrepo2019electron},  dynamic localization~\cite{martinez2008length,tiwari2024dynamical}, Floquet topological states~\cite{minguzzi2022topological,zhan2024perspective}. 
Floquet theory provides a rigorous, systematic, and non-perturbative framework for quantum systems driven by periodic fields for both closed and open systems ~\cite{ivanov2021floquet,kohler2005driven}.  
In non-interaction regime, Floquet-based non-equilibrium green's function (NEGF) method was first developed by H\"anggi et al. based on solving Heisenberg equations of motion for molecular wire setups~\cite{camalet2003current,camalet2004shot}. 
While, the Floquet-based treatments for closed system are well developed ~\cite{shirley1965solution,sambe1973steady}, Floquet methods for open quantum system are diverse~\cite{mori2023floquet,sato2025floquet}, and is still under investigation. In weak system environment coupling regime, Floquet quantum master equation (QME) approaches can offer powerful tools specially for systems in which many-body interactions have to be considered~\cite{lehmann2003laser,mosallanejad2024floquet,mosallanejad2025two}. However, Floquet-based QME can not provide accurate results in strong system environment coupling. In addition, certain two-mode driving protocol such as the STIRAP proven to be very practical in closed systems~\cite{bergmann2019roadmap}. 
Moreover, as topological aspects in strongly driven time-dependent Hamiltonian become a fast-growing and fruitful area of research ~\cite{martin2017topological}, it is imperative to develop Floquet-based NEGF methods that can handle complicated driving protocols.
In the limit of intense driving, having multiple terms with independent frequencies makes the problem fairly complicated.
In this Letter, we first present a two-mode staked (vector-like) Floquet NEGF method for open quantum systems. The time averaged occupation and current formulas are rewritten in terms of two-indexed green's functions. The resulting method has capability to be extended to much sophisticated multi-mode counterpart. We then use our method to study transport of electrons in two-mode driven two-level systems relevant to CDT, highlighting the role of secondary off-diagonal drivings in the current characteristic. 


\textit{Main NEGF equations}.--- 
Often times two primerially objectives are the two-time retarded and advanced Green's function $G^{r,a}(t, t^{\prime})$ obtained from the Kadanoff-Baym Equation (KBE) as
\begin{eqnarray}
\label{eq:1}
\begin{aligned}
&\big(i \partial_t
-h(t)\big)G^{r,a}(t, t^{\prime})
\!-\!
(\Sigma^{r,a} \star G^{r,a})(t, t^{\prime})
\!=\!{I}\delta(t\!-\!t^{\prime}),
\end{aligned}
\end{eqnarray}
where we set $\hbar=1$ and $h(t)$ is the one-body Hamiltonian. The convolution term is denoted as $(\Sigma^{r,a} \star G^{r,a})(t, t^{\prime})=\int \! d t_1 \Sigma^{r,a} $ $(t,t_1) G^{r,a}(t_1, t^{\prime})$, and $\Sigma^{r,a}=\sum_{l} \Sigma^{r,a}_{l}$ are the total retarded and advance self-energies where the index $l$ marks the bath, e.g., $l\in L,R$ in a two terminal setup. ${I}$ is the identity matrix in the Hilbert space ($\mathscr{H}$). 
Throughout this work, all time integrals without explicitly specified limits are taken over $(-\infty,\infty)$.
With regard to the occupation and terminal particle current, the two-time lesser Green's function and current matrix reads
\begin{align}
\MoveEqLeft\!G^{\lessgtr}(t,t^{\prime})\!= \!
\int dt_1 
\int dt_2 
G^{r}(t,t_1)
\Sigma^{\lessgtr}(t_1,t_2) 
G^{a}(t_2,t^{\prime}), \label{eq:2}
\\
\MoveEqLeft\!\operatorname{I}_l(t,t')\!= \!\!\! \int \!\! d t_1 \! 
\big[
G^>\!(t,t_1\!)\Sigma_{l}^{<}\!(t_1,t')
\!+\!G^{<}\!(t,t_1\!)\Sigma_{l}^>\!(t_1,t')
\big]\!,\!\!\label{eq:3}
\end{align}
where $\Sigma^{\lessgtr}=\sum_{l} \Sigma^{\lessgtr}_{l}$. Setting $t'\!=\!t$, the occupation and particle current at the bath $l$ are obtained by $\operatorname{n}(t)\!=\!\operatorname{Tr}(-iG^{<}(t))$ 
and $\operatorname{J}_l(t)\!=\!\operatorname{Tr}(\operatorname{I}_l(t))$, respectively. The two-time structure of above equations makes the time-domain simulation challenging~\cite{tuovinen2023time}. 
The validity of conventional energy-domain NEGF is also restricted to the cases with stationary Hamiltonian and and non-interacting baths. 

\textit{Mixed time-frequency KBE.}---
Without assuming a specific form for $h(t)$, one can define the mixed time-frequency Green's functions via $G^{r,a,\lessgtr}(t,\mathcal{E})=
\int dt' G^{r,a,\lessgtr}(t,t')e^{i\mathcal{E} (t-t')}$. This transformation is founded upon the continuous Fourier transform, thus the energy variable $\mathcal{E}$ is unbound.
The physical meaning of $\mathcal{E}$ may understood as; any particle born in $t'$ could have an independent energy value $\mathcal{E}$. 
Then the KBE for retarded Green's function turns into 
\begin{eqnarray}
\label{eq:4}
\begin{aligned}
\big(\mathcal{E}\!+\!i\partial_t \!-\!h(t)\!\big)G^{r}\!(t, \mathcal{E})
\!-\!\!\!\int\!\!
d\tau \Sigma^{r}(\tau)
e^{ i \mathcal{E}\tau}\!
G^r(t\!-\!\tau,\mathcal{E}) \!=\!{I}\!.~~~
\end{aligned}
\end{eqnarray}
The same EOM holds for $G^a(t,\mathcal{E})$ replacing $\Sigma^{r}$ with $\Sigma^{a}$. Check Appendix for the details of derivation. 
 

\textit{Two-mode V-like Floquet KBE.}--- Consider the Hamiltonian is driven by two separated time-periodic terms characterized by two angular frequencies
$h(t,\omega_1, \omega_2)$. 
Such Hamiltonian is invariant under $t \mapsto t+nT_1+mT_2$ for all $n,m\!\in\!\mathbb{Z}$ where $T_{1,2} = 2\pi/\omega_{1,2}$ are the relevant periods. 
This leads to the same invariance for $G^{r,a}(t, \mathcal{E})$ in the transformed Eq.~(\ref{eq:1}) which in turn allows to employ the 2D Fourier expansion as $G^{r,a}(t, \mathcal{E})\!=\!\sum_{mn}g_{mn}^{r,a}(\mathcal{E})e^{i(n\omega_1+m\omega_2) t}$. 
Motivated by the one-mode Floquet NEGF ~\cite{mosallanejad2024floquet} (see Appendix), one can first obtain a discreet relation for $g_{mn}^{r(a)}(\mathcal{E})$ (see Appendix) and then by specifying a limited integer ranges for $n\in\mathcal{V}_1$ and $m\in\mathcal{V}_2$, it can be casted into an algebraic equation 
\begin{eqnarray}
\label{eq:5}
\begin{aligned} 
\big(\mathbb{E}^F
-H^F-
\Sigma^{r(a)F}(\mathcal{E})\big) 
\mathbf{G_{vF}^{r(a)}}(\mathcal{E})
\!=\!
\operatorname{\mathbf{I}}_{0},
\end{aligned}
\end{eqnarray}
where $\mathbb{E}^F\!=\!\hat{\mathbb{I}}_{\mathcal{V}_2}
\!\otimes\!\hat{\mathbb{I}}_{\mathcal{V}_1}\!\otimes\!\mathcal{E}$. Most importantly $H^F$ is 
\begin{eqnarray}
\label{eq:6}
\begin{aligned}
H^F\!=\!\!\sum_{q,p} 
\!
&\hat{L}_{q,p}
\!\otimes\! 
h_{q,p}
+ 
\hat{N}^{(1)}\!\otimes\!I \omega_{1}
+
\hat{N}^{(2)}\!\otimes\!I\omega_{2},
\end{aligned}
\end{eqnarray}
where $h_{q,p}\!=\!1/(T_1T_2)\!\int_0^{T_1}\!dt_1\!\int_0^{T_1}\!dt_2 h(t_1,t_2)e^{ip\omega_1 t_1}
e^{iq\omega_2 t_2}$.
Here, we defined $\hat{L}_{q,p}\!=\!\hat{L}_{q}^{\mathcal{V}_2}\otimes\hat{L}_{p}^{\mathcal{V}_1}$,
$\hat{N}^{(2)}\!=\!\hat{N}^{\mathcal{V}_2}\!\otimes\!\hat{\mathbb{I}}^{\mathcal{V}_1}$ and $\hat{N}^{(1)}\!=\!\hat{\mathbb{I}}^{\mathcal{V}_2} \otimes \hat{N}^{\mathcal{V}_1}$ 
where
$\hat{L}_{p}^{\mathcal{V}_1}$ ($\hat{L}_{q}^{\mathcal{V}_2}$), $\hat{N}^{\mathcal{V}_1}$ ($\hat{N}^{\mathcal{V}_2}$), and 
$\hat{\mathbb{I}}^{\mathcal{V}_1}$ ($\hat{\mathbb{I}}^{\mathcal{V}_2}$) are the ladder, number, and identity operators/matrices in the integer space $\mathcal{V}_{1} (\mathcal{V}_{2})$.
This $H^F$ is identical to the form of the two-mode Floquet Hamiltonian recently obtained by us in Ref.~\cite{mosallanejad2025two} in the context of the two-mode Floquet QME.
Within two-mode V-like Floquet NEGF, the Floquet retarded (advanced) Green's function is a twofold vertical stack of coefficients as $\mathbf{G_{vF}^{r(a)}}(\mathcal{E})=\sum_{mn} \operatorname{\mathbf{e}}_{m}^{\mathcal{V}_2} \otimes 
\operatorname{\mathbf{e}}_{n}^{\mathcal{V}_1}\otimes  g_{mn}^{r(a)}(\mathcal{E})$ and $\operatorname{\mathbf{I}}_{0}=\operatorname{\mathbf{e}}^{\mathcal{V}_2}_{0}\! \otimes \operatorname{\mathbf{e}}^{\mathcal{V}_{1}}_{0}
\otimes{I}$ where $\operatorname{\mathbf{e}}_{n}^{\mathcal{V}_1 (\mathcal{V}_2)}$ is the $n$th base in unit basis of the space $\mathcal{V}_1$ ($\mathcal{V}_2$).
The Floquet retarded (advanced) self-energy can compactly be given by $\Sigma^{r(a)F}(\mathcal{E})\!=\!\bigoplus_{m\in\mathcal{V}_{2}} \bigoplus_{n\in\mathcal{V}_{1}} \!\Sigma^{r(a)}(\mathcal{E}\!-\!n\omega_1\!-\!m\omega_2)$. 
\vspace{4pt}

\textit{Occupation for the two-mode case.}---  Here, the occupation is a time-dependent observable, $\langle \hat{n}\rangle(t)$. Within the two-mode V-like Floquet NEGF evaluation of $\langle \hat{n}\rangle(t)$ acquires the coefficients $g_{k,l}^{<}(\mathcal{E})$. 
By applying the mixed time-frequency transformation for both sides of Eq.~(\ref{eq:2}), one can show that $G^{\lessgtr}(t,\mathcal{E})$ is invariant under $t \mapsto t+nT_1+mT_2$ just as $G^r(t,\mathcal{E})$ is (see Appendix). This allows us to once again employ the 2D Fourier expansion as $G^{\lessgtr}(t, \mathcal{E})\!=\!\sum_{kl}g_{kl}^{\lessgtr}(\mathcal{E})e^{i(l\omega_1+k\omega_2) t}$ where the coefficients is given by 
\begin{eqnarray}
\label{eq:7}
\begin{aligned}
g_{kl}^{\lessgtr}(\mathcal{E})= \sum_{mn} 
&~g_{k-m~l-n}^{r}(\mathcal{E}_{mn})
\Sigma^{\lessgtr}(\mathcal{E}^{-}_{mn})
g_{mn}^{a}(\mathcal{E}).
\end{aligned}
\end{eqnarray}
Hereafter, we defined the shifted energy as $\mathcal{E}^{-}_{mn}\!=\!\mathcal{E}-n\omega_1-m\omega_2$, see the Appendix for the details of the derivation.
For non-interacting bath and in the energy domain, $\Sigma^{<}$ and $\Sigma^{>}$ have the form:
$\Sigma^{<} (\varepsilon)\!=\!\sum_{l} i f_l(\varepsilon) \, \Gamma_l(\varepsilon)$, and $\Sigma^{>}(\varepsilon)\!=\!\sum_{l}-i\bigl[1-f_l(\varepsilon)\bigr] \, \Gamma_l(\varepsilon)$
where $f_l$ is the Fermi-Dirac distribution for bath $l$ with chemical potential $\mu_l$ and temperature $T_l$, and $\Gamma_l(\varepsilon)$ is the broadening matrix.
Eq.~(\ref{eq:7}) is the key in evaluating the time-average of observables and it can be simplified in a same way for both the commensurate and incommensurate frequencies. 
The time-average of occupation is given by $\langle \hat{n}\rangle\!=\!1/\mathcal{T} \int_0^\mathcal{T} \operatorname{Tr}(-iG^{<}(t)) dt$, where $\mathcal{T}\!=\!LCM(T_1,T_2)$ when the two frequencies are commensurate. However, $\mathcal{T}\!=\!\infty$ when the two frequencies are incommensurate.
For both commensurate and incommensurate scenarios, only the term $g^{<}_{00}(\mathcal{E})\!=\!\sum_{mn} 
(g_{mn}^{a}(\mathcal{E}))^{\dagger}
\Sigma^{<}(\mathcal{E}_{mn})
g_{mn}^{a}(\mathcal{E})$ plays a role in evaluation of $\langle \hat{n}\rangle$. 
Note that, we have employed $g_{-m\,-n}^{r}(\mathcal{E}-n\omega_1-m\omega_2)=g_{mn}^{a}(\mathcal{E})^{\dagger}$, which can be proven based on $G^{a}(t, t^{\prime})\!=\!(G^{r}(t^{\prime},t))^{\dagger}$.
This implies that for numerical evaluation of the time-averaged occupation, $\mathbf{G_{vF}^{a}}(\mathcal{E})$ is preferable to $\mathbf{G_{vF}^{r}}(\mathcal{E})$. 
In summary for the two-mode case, $\langle \hat{n}\rangle=\int (d\mathcal{E}/2\pi)\operatorname{Tr}(-ig_{00}^{<}(\mathcal{E}))$.
Noticeably, Eq.~(\ref{eq:7}) reduces to the single-mode V-like Floquet formulation when the second level indices ($k$ and $m$) are omitted, reversely illustrating its extension to scenarios where the driven modes are more than two. 

\textit{Terminal current for the two-mode case.}--- 
By applying the mixed time-frequency
transformation for both sides of of Eq.~(\ref{eq:3}), one can first arrive at the expression $\operatorname{I}_l(t,\mathcal{E})=G^{>}(t,\mathcal{E})\Sigma_{l}^{<}(\mathcal{E})-G^{<}(t,\mathcal{E})\Sigma_{l}^{>} (\mathcal{E})$ which implies that $\operatorname{I}_l(t,\mathcal{E})$ is also invariant under $t \mapsto t+nT_1+mT_2$ just as the $G^{>}(t,\mathcal{E})$ and $G^{<}(t,\mathcal{E})$ are. Hence, $\operatorname{I}_l(t,\mathcal{E})$ can be expanded by 2D Fourier expansion. 
As $\Sigma_{l}^{<}(\mathcal{E})$ and $\Sigma_{l}^{>}(\mathcal{E})$ are time independent, there is a one-to-one correspondence between the expansion element $\operatorname{I}_{l,mn}(\mathcal{E})$ on LHS and the coefficients $g^{>}_{mn}(\mathcal{E})$ and $g^{<}_{mn}(\mathcal{E})$ on RHS as 
\begin{eqnarray}
\label{eq:8}
\begin{aligned}
\operatorname{I}_{l,mn}(\mathcal{E})=
g_{mn}^{>}(\mathcal{E})\Sigma^{<}_l(\mathcal{E})
-
g_{mn}^{<}(\mathcal{E})\Sigma^{>}_l(\mathcal{E}).
\end{aligned}
\end{eqnarray}
As mentioned, one only needs $\operatorname{I}_{l,00}$ to evaluation the time-averaged of current as: $\langle \operatorname{J}_l\rangle=\int(d\mathcal{E}/2\pi)\operatorname{Tr}(\operatorname{I}_{l,00}(\mathcal{E}))$. Upon substituting
$\Sigma^{<}$ and $\Sigma^{>}$ in $\operatorname{I}_{l,00}$, one finds a Landauer-B\"utakkir-formed expression as 
\begin{eqnarray}
\label{eq:9}
\begin{aligned}
\!\!\langle \operatorname{J}_l\rangle\!=\!\int\!\frac{d\mathcal{E}}{2\pi}
\!\sum_k\!\sum_{mn}\text{T}_{mn}^{lk}(\mathcal{E})
(f_l(\mathcal{E})\!-\!f_k(\mathcal{E}^-_{mn})),
\end{aligned}
\end{eqnarray}
where the transmission coefficients is given by $\text{T}_{mn}^{lk}(\mathcal{E})=\text{Tr}[\Gamma_l(\mathcal{E}) (g_{mn}^{a}(\mathcal{E}))^{\dagger}\,\Gamma_k(\mathcal{E}^-_{mn}) g_{mn}^{a}(\mathcal{E})]$. We note that, our time-average current is not obtained based on specific form of the broadening matrices (e.g. a wire setup) or symmetrization of the current  operator such as the expressions given in the pioneering works~\cite{camalet2003current,camalet2004shot,kohler2005driven}.

\textit{M-mode V-like Floquet NEGF.}---
A driven Hamiltonian can be composed of multiple periodic terms with independent parameters (M-mode)  appearing either as diagonal or off-diagonal matrix elements. In this case, the total Hamiltonian remains invariant under $t \mapsto t+\sum_{\alpha=1}^{M}n_{m}T_{\alpha}$ where $n_{m}\!\in\!\mathbb{Z}$ and $T_{\alpha} = 2\pi/\omega_{\alpha}$. 
Therefore, the mixed time-frequency KBE can be transformed into a multi-mode discreet Floquet KBE with extending the number of indices in the expansion elements as $g_{n_M,...,n_1}^{r(a)}(\mathcal{E})$ where each index $n_{\alpha}$ runs over a limited symmetric integers each forms a separate space denoted by $\mathcal{V}_{\alpha}$. 
While the resulting truncated multi-mode Floquet KBE retain the same algebraic structure shown in Eq.~(\ref{eq:5}), its components, in particular the Floquet Hamiltonian, must be generalized based on multi-dimensional complex Fourier series. Floquet energy variable defines as $\mathbb{E}^F\!=\!\hat{\mathbb{I}}_{tot}
\!\otimes\!
\mathcal{E}$
where 
$\hat{\mathbb{I}}_{tot}=\hat{\mathbb{I}}^{\mathcal{V}_M}
\!\otimes\!
\cdots
\!\otimes\!
\hat{\mathbb{I}}^{\mathcal{V}_1}$. 
The Floquet Hamiltonian becomes 

\begin{eqnarray}
\label{eq:10}
\begin{aligned}
&H^F\!=\!\! \sum_{q_M,\cdots,q_1} 
\!\!\hat{L}_{q_M,\cdots,q_1}
\!\otimes\! 
h_{q_M,\cdots,q_1}
+\!\sum_{\alpha}\!
\hat{N}^{(\alpha)}\!\otimes\!I\omega_{\alpha},
\end{aligned}
\end{eqnarray}
where 
$h_{q_M,\cdots,q_1}\!=\!1/(T_1 \cdots T_M\!)\!\int_0^{T_1}\!dt_{1} 
\cdots\!\int_0^{T_M}\!dt_M h(t_1,
\cdots\!$
$,t_M) e^{iq_1\omega_1 t_1} \cdots e^{iq_M\omega_M t_M}$.
Here, we redefined multi-mode ladder and number operators as $\hat{L}_{q_M,\cdots,q_1}$ $\!=\!\hat{L}_{q_M}^{\mathcal{V}_M}\!\otimes \cdots \otimes\!\hat{L}_{q_1}^{\mathcal{V}_1}$,
and $\hat{N}^{(\alpha)} = \big(\bigotimes_{j=M}^{\alpha+1} \hat{\mathbb{I}}^{\mathcal{V}_j} \big) 
\otimes 
\hat{N}^{\mathcal{V}_\alpha} \otimes \big(\bigotimes_{j=\alpha-1}^{1} \hat{\mathbb{I}}^{\mathcal{V}_j} \big)$.  
The multi-mode Floquet retarded (advanced) self-energies is redefine as $\Sigma^{r(a)F}(\mathcal{E})\!=\!\bigoplus_{n_M\in\mathcal{V}_{M}} \cdots \bigoplus_{n_1\in\mathcal{V}_{1}} \!\Sigma^{r(a)}(\mathcal{E}\!-\!n_1\omega_1\!-\!\cdots\!-\!n_M\omega_M)$. 
Lastly, multi-mode V-like Floquet retarded (advanced) Green's function is a multi-fold vertical stack of coefficients defined as $\mathbf{G_{vF}^{r(a)}}(\mathcal{E})\!=\!\sum_{n_M \cdots n_1} \operatorname{\mathbf{e}}_{n_M}^{\mathcal{V}_M} \otimes \cdots \otimes 
\operatorname{\mathbf{e}}_{n_1}^{\mathcal{V}_1} \otimes g_{n_M, \cdots,n_1}^{r(a)}(\mathcal{E})$ and $\operatorname{\mathbf{I}}_{0}=\operatorname{\mathbf{e}}^{\mathcal{V}_M}_{0} \otimes \cdots \otimes \operatorname{\mathbf{e}}^{\mathcal{V}_{1}}_{0}
\otimes I$. 
Hereafter, we identify the vector of indices arranged in the decedent way as $\mathbf{n}_{\downarrow}\equiv\{n_M, \cdots,n_1\}$ and the vector of driven frequencies as $\mathbf{\omega}_{\downarrow}\equiv\{\omega_M, \cdots,\omega_1\}$ to compactly define multi-mode shifted energy variable as $\mathcal{E}^-_{\mathbf{n}_{\downarrow}}\equiv\mathcal{E}\!-\!\mathbf{n}_{\downarrow}\cdot\mathbf{\omega}_{\downarrow}$.
With regards to the occupation, the main structure of Eq.~(\ref{eq:7}) remains unchanged. However, besides  modification of the shifted energy variable $\mathcal{E}_{mn} \mapsto \mathcal{E}_{\mathbf{n}_{\downarrow}}$, one should update indices of the retarded and advanced green's functions (e.g., $g_{m,n}^{r}(\mathcal{E})\mapsto g_{n_M,...,n_1}^{r}(\mathcal{E})\equiv g_{\mathbf{n}_{\downarrow}}^{r}(\mathcal{E})$). Nonetheless, the time-average of occupation is given by 
\begin{eqnarray}
\label{eq:11}
\begin{aligned}
\langle \hat{n}\rangle\!=\!\!\int \! (d\mathcal{E}/2\pi)
\operatorname{Tr}
\big(
\sum_{\mathbf{n}_{\downarrow}} 
g_{\mathbf{n}_{\downarrow}}^{a}(\mathcal{E})^{\dagger}
\Sigma^{<}(\mathcal{E}^-_{\mathbf{n}_{\downarrow}})
g_{\mathbf{n}_{\downarrow}}^{a}(\mathcal{E})
\big).
\end{aligned}
\end{eqnarray}
Following same path, the modified Eq.~(\ref{eq:9}) gives the time-averaged of terminal particle current as  
\begin{eqnarray}
\label{eq:12}
\begin{aligned}
\!\!\langle \operatorname{J}_l\rangle\!=\!\!\int\!\frac{d\mathcal{E}}{2\pi}
\!\!\sum_k\sum_{{\mathbf{n}_{\downarrow}}}\!
\text{T}_{{\mathbf{n}_{\downarrow}}}^{lk}\!(\mathcal{E})
(f_l(\mathcal{E})\!-\!f_k(\mathcal{E}^-_{\mathbf{n}_{\downarrow}})),
\end{aligned}
\end{eqnarray}
with the transmission coefficients is given by $\text{T}_{\mathbf{n}_{\downarrow}}^{lk}(\mathcal{E})=\text{Tr}[\Gamma_l(\mathcal{E}) (g_{\mathbf{n}_{\downarrow}}^{a}(\mathcal{E}))^{\dagger}\,\Gamma_k(\mathcal{E}^-_{\mathbf{n}_{\downarrow}}) g_{\mathbf{n}_{\downarrow}}^{a}(\mathcal{E})]$.

\begin{figure}
	\begin{center}
\includegraphics[width=8.6cm]{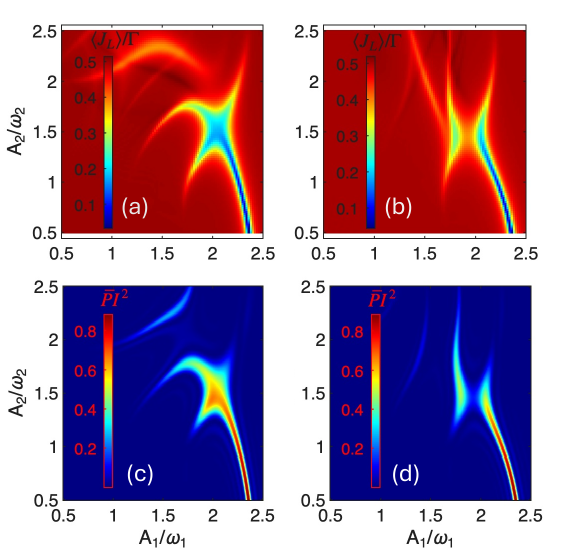}
\end{center}\caption{\label{fig:2} Map of two-mode coherent destruction of current flow in a two-level system connected to the left and right terminals (a) for $\omega_2=0.4825\,\omega_{1}$ and (b) for $\omega_2=0.5175\,\omega_{1}$. (c) and (d) Corresponding maps of localization, obtained for the closed system.}
\end{figure}

\textit{Two-mode current suppression}---
As a tractable yet nontrivial application, we consider the simplest CDT in a two-level system (TLS), in which the onsite energies are modulated by $\pm f_1(t)=(\pm A_1/2) cos(\omega_1 t)$, while levels are coupled via the nearest neighbor hopping energy $\Delta$. 
\begin{figure}[h]
	\begin{center}
	\includegraphics[width=8.6cm]{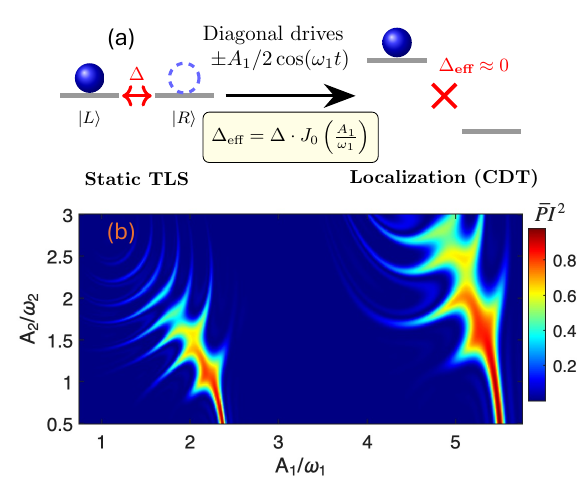}
	\end{center}\caption{\label{fig:1} (a) Schematic of a single mode, diagonal, driven coherent destruction of tunneling (CDT) in a two-level system (TLS). (b) Map of localization, square of the population imbalance $\bar{PI}^2$, in presence of sinonasal off-diagonal drive.}
\end{figure}
Within the single mode CDT (original) problem, tunneling suppressed on the zeros of $J_0(A_1/\omega_1)$, $\{\text{2.4},\text{5.5},...\}$, see the schematics in Fig.~(\ref{fig:1}) (a). Our modified system is composed of the conventional CDT plus a secondary off-diagonals periodic term, $f_2(t)=(A_2/2) sin(\omega_2 t)$, such that matrix Hamiltonian reads
\begin{eqnarray}
\label{eq:13}
h(t)=
\begin{bmatrix}
\epsilon_L+f_1(\omega_1,t) &\! \Delta+f_2(\omega_2, t) \\
\Delta+f_2^*(\omega_2, t) &\! \epsilon_R-f_1(\omega_1,t)  
\end{bmatrix}.
\end{eqnarray}
In the closed system, one can initializing the particle in one of the sites and numerically solve the time-dependent schr\"odinger equation, evaluate the long-time average of populations, $\bar{P}_{\alpha}\!=\!\tau^{-1}\!\int_0^\tau \! dt P_{\alpha}(t)$, and take the $\bar{PI}^2=(\bar{P}_{L}-\bar{P}_{R})^2$ as a measure for tunneling suppression (particle localization). Interestingly, in the presence of the second driving term, $f_2(t)$, and with choosing $\omega_2$ around $\omega_1/2$, the particle localization shows interesting patterns as function of $A_1$ and $A_2$ (the map of localization)as shown in Fig.~(\ref{fig:1}) (b). Such calculation in a closed system serves as a guiding tool for the open system. 
To obtain Fig.~(\ref{fig:1}) (b), we set, $\epsilon_{L,R}=0$, $\Delta=0.1$, $\omega_1=20\Delta$, as CDT is pronounced in the high frequency regime, $\omega_2=0.4925\,\omega_1$, and plot the map of localization.
One expects that under high electrochemical difference in an open system, which is our main focus, the current flow mimics the particle localization of the driven closed system. Although, employing two-frequency driving (1:2 ratio) on the diagonal are considered~\cite{farrelly1993two}, to the best of our knowledge such map of localization has not be reported yet. We found the distortion of localization lines is ultra sensitive to the value of $\omega_2$ such that a small changes around $\omega_2/2$ will drastically modify the pattern of localization. 
Localization can also be investigated by entropy, $S(t)=-\sum_\alpha \bar{P}_{\alpha}(t) \text{ln}(\bar{P}_{\alpha}(t)) $. However, the long-time average of entropy is also deliver similar localization map. 
To show the capability of the multi-mode Floquet NEGF, we set $\omega_1=10\Delta$, $\mu_{L(R)}=+(-)35\Delta$, and obtain the left-terminal current, $\langle\operatorname{J}_L\rangle$, at the second frequencies $\omega_2=0.4825\,\omega_{1}$ and $\omega_2=0.5175\,\omega_{1}$,
for a range of $A_1$ and $A_2$, as shown in the Figs.~(\ref{fig:2}) (a) and (b). Here, the level $\ket{L}$ ($\ket{R}$) is connected only to a left (right) terminal via a small coupling $\Gamma_{L(R)}=\Delta/40$ at very low temperature, $4.2 \text{K}$, to produce a narrow dip in current flow (notice that 1D current curves in Figs.~(\ref{fig:2}) (a) and (b) exhibit a dip around $A_1/\omega_1=2.4$ at low $A_2/\omega_2$). Convergence has achieved by setting the $N_{1,2}=7$, which determines the integer spaces $\mathcal{V}_{1,2}$. The star like pattern in Fig.~(\ref{fig:2}) (a) and the splitted lines in Fig.~(\ref{fig:2}) (b) are identical to the  localization patterns shown in Figs.~(\ref{fig:2}) (c) and (d). 
Finally we reports two rational observations without explicitly showing the relevant plots. First the effects of having higher coupling $\Gamma$ is to make the dip patterns wider. 
Secondly, the pattern of current suppression does not show sensitivity to the phase difference between drivings possibly due to the coupling destroy phase coherence.   
 
\textit{Conclusions}--- 
We have derived the multi-mode Floquet NEGF approach, which enables quantum transport studies of systems under multiple time-periodic driving terms. We have shown how to obtain the time-averaged population and current under such drivings. Guided by dynamical localization in a two-level system driven by two independent periodic terms ---one diagonal and one off-diagonal--- we explored how the secondary driving alters the dynamical obstruction of current flow in a molecular wire setup. The main finding is that, with appropriate choice of amplitude and frequency for the secondary off-diagonal drive, one can achieve higher control over the suppression of current flow. By circumventing real-time NEGF, this method allows us to explore richer physics of multi-chromatically driven open systems. The multi-mode Floquet NEGF may also apply to time-dependent topological transport problems such as Thouless pumping. 

\textit{Acknowledgments}---W. D. acknowledges the support from National Natural Science Foundation of China (No. 22361142829 and No. 22273075) and Zhejiang Provincial Natural Science Foundation (No.XHD24B0301). V. M. acknowledges the funding from the Summer Academy Program for International Young Scientists(Grant No. GZWZ[2022]019).

\bibliography{bib_MMVFNEGF}
\appendix

\begin{onecolumngrid} 
\centering 
\section*{END MATTER} 
\end{onecolumngrid}

\textit{Appendix: Conventional NEGF.}--- To justify our arguments on conventional NEGF, in the main context, we start recapping the common scenario in which the Hamiltonian is time-independent (conventional NEGF). Most often terminals are treated as non-interacting metallic baths, resulting in the self-energies being solely functions of the time difference $\tau\!=\!t-t^{\prime}$, as $\Sigma^{r,a,\lessgtr}(t, t^{\prime})=\Sigma^{r,a,\lessgtr}(\tau)$. For these baths, all Green's functions exhibit the time-translation invariance, which means $G^{r,a,\lessgtr}(t, t^{\prime})\!=\!G^{r,a,\lessgtr}(\tau)$. 
This further allows us to express the KBE purely in terms of time differences as $\big(i {d}/{d \tau}-h\big)G^{r,a}(\tau)\!-\!\!\int \! d \tau_1 \Sigma^{r,a}(\tau\!-\!\tau_1) G^{r,a}(\tau)\!=\!{I} \delta(\tau)$, which in turn results in a significant simplification of the KBE in the energy domain after performing the continuous Fourier transformation with respect to $\tau$ as: $\big(E\!-\!h\!-\!\Sigma^{r,a}(E) \big) G^{r,a}(E)\!=\!I$. 
Additionally, we can first arrive at the lesser green's function in terms of $\tau$, as in $G^{\lessgtr}(\tau)\!=\!\int\! d \tau_1\!\int\! d \tau_2G^r(\tau-\tau_1) \Sigma^{\lessgtr}(\tau_1-\tau_2) G^a(\tau_2)$, where we defined $\tau_{1,2}=t_{1,2}-t^{\prime}$. Then, applying the Fourier transformation with respect to $\tau$ gives the Fourier spectrum $G^{\lessgtr}(E)\!=\!G^r(E)\Sigma^{\lessgtr}(E)G^a(E)$, with $\Sigma^{<}(E)\!=\!\sum_{l} f_l(E)[\Sigma^{r}_{l}(E)-\Sigma^{a}_{l}(E)]$ for the non-interaction bath. 
Furthermore, electron number (occupation) is defined as $\langle \hat{n}\rangle\!=\!\operatorname{Tr}(-iG^{<}(\tau\!=\!0))\!=\!\int 
(dE/2\pi)\operatorname{Tr}(-iG^{<}(E))$. 
The same procedures can be implemented for Eq. (\ref{eq:3}), to derive the Fourier spectrum for the current matrix $\operatorname{I}_{l}(E)\!=\!G^{>}(E)\Sigma_l^{<}(E)+G^{<}(E)\Sigma_l^{>}(E)$, and then to evaluate the terminal current (as an observable) 
$\langle \operatorname{J}_{l}\rangle\!=\!\operatorname{Tr}\big(
\operatorname{I}_{l}(\tau\!=\!0)\big)\!=\!\int (dE/2\pi) \operatorname{Tr}\big(\operatorname{I}_{l}(E)\big)$. Also, Landauer-B\"utakkir-form can be extracted by simplifying $\operatorname{I}_{l}(E)$. 
All these simplifications in the energy domain essentially occur because Hamiltonian is time-independent and the Fourier transformation of the the time domain convolution is the product in the energy domain.
\vspace{4pt}

\textit{Details of deriving the mixed time-frequency KBE.}
Performing the mixed time-frequency transformation, the first part on the left of the KBE becomes: $\big(\mathcal{E}+i \partial_t-h(t)\big)G^r(t, \mathcal{E})$. The convolution part of the KBE simplifies first to $\int d t_1 \Sigma^r(t-t_1)e^{i \mathcal{E}(t-t_1)} G^r(t_1,\mathcal{E})$ and then, using the convolution shift property, it becomes $\int d\tau \Sigma^r(\tau)e^{ i \mathcal{E}\tau}G^r(t-\tau, \mathcal{E})$, where we redefine the time difference as $\tau\!=\!t-t_1$. The right side of the KBE becomes ${I}$.
\vspace{4pt}

\textit{One-mode V-like Floquet NEGF.}---
Here, we briefly recap how one can obtain an expression for Floquet KBE when the Hamiltonian has one periodic term, $h(t+T)\!=\!h(t)$. Here, the Hamiltonian may parametrize by a single frequency as: $h(t,\omega)$, where $\omega\!=\!2\pi/T$. Under this assumption, the two-time Green's functions given in Eqs.~(\ref{eq:1}), features discrete two-time translation symmetry meaning $G^{r, a}(t\!+\!T, t^{\prime}\!+\!T)\!=\!G^{r, a}(t,t^{\prime})$. Thus, the mixed time-frequency Green's function features the discrete time translation symmetry as $G^{r, a}(t, \mathcal{E})=G^{r, a}(t+T, \mathcal{E})$. 
We note that discrete two-time translation symmetry also emerges when the self-energy is periodic, and the Hamiltonian is static. However, the analysis of these scenarios is reserved for future study.
As $G^{r,a}(t, \mathcal{E})$ is periodic in $t$, it can be expanded by discrete Fourier expansion as $G^{r,a}(t, \mathcal{E})\!=\!\sum_{m}g_{m}^{r,a}(\mathcal{E})e^{im\omega t}$, $m \in \mathbb{Z}$. 
This essentially enables us to perform the integration over $\tau$ in Eq.~(\ref{eq:4}). 
Redefining the self-energy in the energy domain by $\Sigma^r(\mathcal{E}-m\omega)\!=\!\int d\tau \Sigma^r(\tau)
e^{ i (\mathcal{E}-m\omega)\tau}$, we arrive at the following expression for Eq.~(\ref{eq:4}) 
\begin{eqnarray}
\label{eq:app1}
\begin{aligned} 
\sum_{m}
\!\big(\mathcal{E}
\!-\!h(t)-m\omega\!-\!
\Sigma^r(\mathcal{E}-m\omega)\big) 
g_{m}^{r}(\mathcal{E})e^{im\omega t}=I.~~~~~
\end{aligned}
\end{eqnarray}
The component $g_{m}^{r}(\mathcal{E})$ is obtained by multiplying Eq.~(\ref{eq:app1}) by 
$e^{-in\omega t}$ and averaging over one period, 
$1/T \int_0^T () dt$ as
\begin{eqnarray}
\label{eq:app2}
\begin{aligned} 
\sum_{m}
\!\big(
\!-\!h_{m-n}\!+\!
\big(\mathcal{E}
\!-\!m\omega\!-\!
\Sigma^r(\mathcal{E}\!-\!m\omega)
\big)\delta_{nm}
\big)
g_{m}^{r}(\mathcal{E})={I}\delta_{n0},~~~
\end{aligned}
\end{eqnarray}
where $h_{m-n}=1/T \int_0^T h(t)e^{i(m-n)\omega t}dt$. Running $(m,n)$ over a limited integer, $-N \leq n,m \leq N$ where $N$ is a positive integer, one can express the truncated version of Eq.~(\ref{eq:app2}) with $\big(\mathbb{E}^F-H^F-\Sigma^{rF}(\mathcal{E})\big) \mathbf{G_{vF}^r}(\mathcal{E})=\operatorname{\mathbf{I}}_{0}$ (Eq.~(\ref{eq:5}) in main text) where $\mathbb{E}^F=\hat{\mathbb{I}}\otimes\mathcal{E}$, with $\hat{\mathbb{I}}$ being the identity matrix in truncated integer space, denoted by $\mathcal{V}$. 
Floquet Hamiltonian $H^F$ is defined by its matrix block elements $(H^F)_{nm}=h_{m-n}\!+\!{I} m\omega\delta_{nm}$. Because the block elements of $H^F$ depend only on the Fourier index difference d=m-n, $H^F$ can be compactly expressed as
\begin{eqnarray}
\label{eq:app13}
\begin{aligned}
H^F=\sum_{d} \hat{L}_{d}\otimes h^{d}+\hat{N}\otimes I\omega
\end{aligned}
\end{eqnarray}
where $\hat{L}_{d}$ are ladder operators (sparse matrices with 1 on the $d$th off-diagonal), and $\hat{N}$ (the number operator) is the diagonal matrix with elements $N_{nm}=m\delta_{nm}$.
We refer to $\Sigma^{rF}(\mathcal{E})=\bigoplus_{n} \Sigma^r(\mathcal{E}-n\omega)$ as the Floquet retarded self-energy with block elements $(\Sigma^{rF}(\mathcal{E}))_{nm}= 
\Sigma^r(\mathcal{E}\!-\!m\omega)\delta_{nm}$. 
Within the one-mode case,  $\mathbf{G_{vF}^r}(\mathcal{E})=$ $\sum_m \operatorname{\mathbf{e}}_{m}^{\mathcal{V}} \otimes g_{m}^{r}(\mathcal{E})$ is a vertical/vector (V-like) stack of coefficient matrices $g_{m}^{r}(\mathcal{E})$ and $\operatorname{\bf{I}}_{0}=\operatorname{\mathbf{e}}_{0}^{\mathcal{V}}\otimes{I}$. Here, $\operatorname{\mathbf{e}}_{m}^{\mathcal{V}}$ is equivalent to the $\operatorname{\mathbf{e}}_{m+N+1}$ base in the standard basis vector of $\mathbb{R}^{2N+1}$. 
In summary, the two most essential keys of the V-like Floquet KBE are the definition of $G^{r,a}(t,\mathcal{E})$ and its invariance under $t \mapsto t+nT$.
\vspace{4pt}

\textit{Discreet equation for $g^{r,a}_{nm}$.}---
Relaying on the single frequency Floquet~\cite{mosallanejad2024floquet}, one can realize that the two-time Green's functions, given by Eqs.~(\ref{eq:1}), feature the discrete two-time translation symmetry, $G^{r, a}(t\!+\!nT_1+mT_2,t^{\prime}\!+\!nT_1+mT_2)\!=\!G^{r, a}(t,t^{\prime})$, which implies that the mixed time-frequency Green’s functions, $G^{r, a}(t,\mathcal{E})$ also feature discrete time translation symmetry meaning $G^{r, a}(t\!+\!nT_1+mT_2, \mathcal{E})=G^{r,a}(t,\mathcal{E})$. This property allows us to employ the 2D Fourier expansion for $G^{r, a}(t, \mathcal{E})$. 
To robustly derive an equation similar to Eq.~(\ref{eq:app2}), we temporarily replace the $t$ in Hamiltonian with $t_1$ and $t_2$, based on the two driving terms associated with $\omega_1$ and $\omega_2$, apply the chain rule $\partial_t=\partial_{t_1}+\partial_{t_2}$, and then employ the expansion $G^{r,a}(t_1,t_2,\mathcal{E})\!=\!\sum_{m_1m_2}g_{m_1m_2}^{r,a}(\mathcal{E})e^{im_1\omega_1 t_1} e^{im_2\omega_2 t_2}$.
Then, Eq.~(\ref{eq:5}) and its advanced counterpart can be expressed as: $\sum_{m_1m2}
\!\big(\mathcal{E}\!-\!h(t_1,t_2)-m_1\omega_1-m_2\omega_2\!-\!\Sigma^{r,a}(\mathcal{E}-m_1\omega_1-m_2\omega_2)\big) g_{m_1m_2}^{r,a}(\mathcal{E})e^{im_1\omega_1 t_1}e^{im_2\omega_2 t_2}={I}$. 
Multiplying both sides by $e^{-in_1\omega_1 t_1}e^{-in_2\omega_2 t_2}$, and taking the double average of one-period, $1/(T_1 T_2)\int_0^{T_1} dt_1\int_0^{T_2} dt_2$ tuns Eq.~(\ref{eq:4}) into 
\begin{eqnarray}
\label{eq:app4}
\begin{aligned} 
&\sum_{m_1 m_2}
\!\big(-\!
h_{m_1-n_1,m_2-n_2}\!+\!
\big(\mathcal{E}
\!-\!m_1\omega_1\!-\!
\!m_2\omega_2\!-\!
\Sigma^{r,a}(\mathcal{E}
-\!
m_1\omega_1\!-\!m_2\omega_2)\big) \delta_{n_2m_2}
\delta_{n_1m_1}
\big) 
g_{m_1,m_2}^{r,a}(\mathcal{E})\!=\!{I}\delta_{n_10}\delta_{n_20},~~~
\end{aligned}
\end{eqnarray}
where $h_{m_1-n_1,m_2-n_2}=1/(T_1T_2) \int_0^{T_1} dt_1 \int_0^{T_1} dt_2 h(t_1,t_2)e^{i(m_1-n_1)\omega_1 t_1}
e^{i(m_2-n_2)\omega_2 t_2}$. As the difference between $m_i$ and $n_i$ matters, the block elements $h_{m_1-n_1,m_2-n_2}$ can be expressed only by two indices as $h_{d_1,d_2}$ where $d_i=m_i-n_i$. 
\vspace{4pt}

\textit{Casting Discreet Floquet KBE into an algebraic equation.}---
The two sets of indices in Eq.~(\ref{eq:app4}), $n_{1,2}$ and $m_{1,2}$, should run over limited integers, $-N_i \leq n_i,m_i \leq N_i$ where $i\in 1,2$. 
Focusing on $\operatorname{I}\delta_{n_10}\delta_{n_20} $, it is natural to fix on the pair $(n_1,n_2)$, and sweep over $m_1$ and $m_2$, respectively. We can then form a grand matrix algebraic equation whose rows and columns are determined by the pairs $(n_1,n_2)$ and $(m_1,m_2)$, respectively. 
If we first run over $n_1$ (inner loop) and then $n_2$ (outer loop), the right side of Eq.~(\ref{eq:app4}) becomes $\operatorname{\mathbf{I}}_{0}=\operatorname{\mathbf{e}}^{\mathcal{V}_{2}}_{0}
\otimes
\operatorname{\mathbf{e}}^{\mathcal{V}_{1}}_{0}
\otimes{I}$. 
Additionally, sweeping first on $m_1$ and then $m_2$ indicates double stacking as $\mathbf{G_{v}^{rF}}(\mathcal{E})=$ $\sum_{m_2 m_1} \operatorname{\mathbf{e}}_{m_2}^{\mathcal{V}_2} \otimes 
\operatorname{\mathbf{e}}_{m_1}^{\mathcal{V}_1} \otimes
g_{m_1m_2}^{r}(\mathcal{E})$.
Setting $t_{1,2}=t$ and re-indexing ($m_2 \to m$, $m_1 \to n$) in the expansion of $G^{r,a}(t_1,t_2,\mathcal{E})$, one obtain $G^{r,a}(t,t^{\prime})=\sum_{m,n}\int (d\mathcal{E}/2\pi)g_{m,n}^{r,a}(\mathcal{E})
e^{-i(\mathcal{E}-n\omega_1-m\omega_2)t} 
e^{i\mathcal{E}t^{\prime}}$.
\vspace{4pt}

\textit{Mixed time-frequency lesser Green's function.}---
Applying the mixed time-frequency transformation for both sides of Eq.~(\ref{eq:2}) while taking $t_2\!=\!t_1\!-\!\tau\!$, one can first obtain the general expression $G^{<}(t,\mathcal{E})\!=\!\int \! dt_1 G^{r}(t,t_1)e^{i\mathcal{E}(t-t_1)}\!\int d\tau \Sigma^{<}(\tau)e^{i\mathcal{E}\tau}G^{a}(t_1-\tau$ $,\mathcal{E})$, for which integration over $\tau$ can only be performed when the time invariance property of $G^a(t_1-\tau$ $,\mathcal{E})$ is determined. 
For the two-mode case, we shall employ $G^{a}(t_1-\tau,\mathcal{E})=\sum_{m,n}\int g_{m,n}^{a}(\mathcal{E})
e^{i(n\omega_1+m\omega_2)t_1}e^{-i(n\omega_1+m\omega_2)\tau}$ which results in $G^{\lessgtr}(t,\mathcal{E})\!=\!\sum_{m,n}e^{i(n\omega_1+m\omega_2)t}\int dt_1G^{r}(t,t_1)e^{i\mathcal{E}(t-t_1)}e^{-i(n\omega_1+m\omega_2)(t-t_1)}
\Sigma^{\lessgtr}(\mathcal{E}-n\omega_1-m\omega_2)g_{m,n}^{a}(\mathcal{E})$. 
Performing the integration over $t_1$ gives 
\begin{eqnarray}
\label{eq:app5}
\begin{aligned} 
G^{\lessgtr}(t,\mathcal{E})\!=\!\sum_{m,n} e^{i(n\omega_1+m\omega_2)t}
G^{r}(t,\mathcal{E}-n\omega_1-m\omega_2)
\Sigma^{\lessgtr}(\mathcal{E}-n\omega_1-m\omega_2)
g_{m,n}^{a}(\mathcal{E}), 
\end{aligned}
\end{eqnarray}
which indicates that discreet time invariance property of $G^{\lessgtr}(t,\mathcal{E})$ follows the invariance property of $G^r(t,\mathcal{E})$.
Here, we can clearly observe that the same shifts in $\mathcal{E}$ only occur for $G^{r}$ and $\Sigma^{\lessgtr}$. 
In the final step, we perform 2D Fourier transformation of Eq.~(\ref{eq:app5}). This requires temporarily replacing $t$ with $t_1$ and $t_2$ in expanding $G^{\lessgtr}(t,\mathcal{E})$ and $e^{i(n\omega_1+m\omega_2)t}G^{r}(t,\mathcal{E}-n\omega_1-m\omega_2)$ based on their associations with $\omega_1$ and $\omega_2$, as we did in deriving Eq.~(\ref{eq:app4}). Taking the double average, $1/(T_1 T_2)\int_0^{T_1}dt_1\int_0^{T_2} dt_2$ results in Eq.~(\ref{eq:7}).
\end{document}